\documentclass[authoryear,12pt,5p,times]{elsarticle}
\usepackage{graphicx}
\usepackage{color}
\usepackage{amssymb}

\usepackage{txfonts}
\journal{Astronomy and Computing}
\begin{document}
%
\begin{frontmatter}
\title{Cosmological calculations on the GPU}

\author[kav]{D. Bard}
\ead{djbard@slac.stanford.edu}
\author[sie,niu]{M. Bellis\fnref{fn1}}
\ead{mbellis@siena.edu}
\author[sta]{M. Allen}
\author[foo]{H. Yepremyan}
\author[mia]{J. Kratochvil}

\address[kav]{Kavli Institute for Particle Astrophysics and Cosmology, Stanford, CA 94309}
\address[sie]{Department of Physics and Astronomy, Siena College, Loudonville, NY, 12211}
\address[niu]{Department of Physics, Northern Illinois University, DeKalb, IL, 60115}
\address[sta]{Department of Physics, Stanford University, Stanford, CA, 94309}
\address[foo]{Foothill College, Los Altos Hills, CA, 94022}
\address[mia]{Department of Physics, University of Miami, Florida, FL, 33143}

\fntext[fn1]{Currently at Siena College.}


\begin{abstract}
  Cosmological measurements  require the calculation of nontrivial quantities over  large datasets.
      The next generation of survey telescopes  will yield measurements of billions of galaxies. The scale of these datasets, and the nature of the calculations involved, make cosmological calculations ideal models for implementation on graphics processing units (GPUs).
  We consider two cosmological calculations, the two-point angular correlation function and the aperture mass statistic, and aim to improve the calculation time by constructing code for calculating them on the GPU.
  Using CUDA, we implement the two algorithms on the GPU and compare the calculation speeds to comparable code run on the CPU.
  We obtain a code speed-up of between 10 - 180x faster, compared to performing the same calculation on the CPU. The code has been made publicly available.
  GPUs are a useful tool for cosmological calculations, even for datasets the size of current surveys, allowing calculations to be made one or two orders of magnitude faster.
  \end{abstract}

\end{frontmatter}

%

\section{Introduction}
\label{sec:intro}

The use of graphics processing units (GPUs) in scientific computing has been steadily growing in fields as diverse as bioinformatics, QCD lattice calculations and seismology (see, for example,~\cite{bio, qcd, seis}). 
In astronomy, GPUs have proven useful in many different  computationally intensive problems such as N-body simulations (\cite{2012JCoPh.231.2825B, nitadori}) and radio astronomy measurements (\cite{clark}). 
GPU techniques have succeeded in reducing compute times for these difficult calculations by up to a factor of 100, and in this work we show that a similar reduction can be achieved for the calculation of cosmological quantities. 
 
The next generation of large-scale astronomical surveys (such as the Dark Energy Survey~\footnote{http://www.darkenergysurvey.org}, PanSTARRS~\footnote{http://pan-starrs.ifa.hawaii.edu/}, and  the Large Synoptic Survey Telescope~\footnote{www.lsst.org}) will produce enormous amounts of data, with measurements of billions of stars and galaxies. 
The problems associated with processing such a large volume of image data and how to structure and access a database containing tens of petabytes of information has been studied and discussed at length in the literature (e.g. ~\cite{way, berriman, brunner}). 
In this paper, we address the challenges an astronomer will face attempting to analyze this information, once it has been obtained from a central database. 
With such a large volume of data the statistical uncertainties on many cosmological quantities will be reduced by orders of magnitude compared to present limits, but in order to take advantage of this data new computation methods must be developed. 
Methods used today to calculate cosmological quantities tend to be of complexity $O(n^{\ 2})$ (where $n$ is the number of data points), which is not computationally feasible with the billions of measurements expected from future surveys, even with the expected improvements in computer hardware. 

Many cosmological calculations require independent calculations of the same quantity for all data points, which makes them ideal candidates for parallelization (where each calculation is farmed out to a different processing device). 
In the past this has often been handled by using many CPUs in a computing cluster environment, but building these clusters or buying time on these systems can be expensive for researchers. 
GPU computing has brought a significant amount of this computational power to the desktop and is therefore more affordable for individual analysts. 
As GPU computing develops and becomes more widely used by the broader astronomy community, we foresee performing calculations in the near future that are currently computing-limited.
In this paper, we describe the GPU implementation of two of these cosmological 
calculations, the two-point angular correlation function and aperture mass statistic, 
using the CUDA programming language (\cite{cuda}). 

Large-scale structure in the universe is a valuable probe of the composition and evolution of matter in the universe, and  can be used to constrain models of cosmology. 
Galaxies are good tracers of the total matter in the universe; although we can only directly detect the luminous matter, galaxies form around concentrations of dark matter. 
We can therefore characterize large scale structure of matter in the universe using the
clustering of galaxies on different length scales, which is measured using the angular correlation
function. The two-point angular correlation function (or matter power spectrum in Fourier space) is based on galaxy number counts, and measures the excess or depletion of pairs of galaxies as a function of separation, compared to a random distribution. 
At small scales  ($\approx 100 h^{-1}Mpc$) we can measure the imprint of the baryon acoustic oscillations, which gives information on the phases of acoustic waves at recombination. 
At larger scales the matter power spectrum has not been affected by radiation or baryons, so it is a rare probe of primordial fluctuations and inflation.  
See~\cite{bassett} for a full review of the subject, \cite{peebles} for a description of large-scale structure in the universe and~\cite{cole, eisenstein} for the first measurements of baryon acoustic oscillations. 
Calculating a correlation function over billions of galaxies, and at separations ranging from arcseconds to degrees, requires significant computational power, which scales with the square of the number of galaxies. 
As such it is an excellent candidate for implementation on the GPU.  
Higher order correlation functions (such as the three-point correlation function) can also be used to probe non-Gaussian features in the galaxy distribution. 
These calculations are even more computationally expensive than the two-point correlation function, and would potentially benefit enormously from parallelization on the GPU.

As we were preparing this paper, we became aware of work by~\cite{ponce}  
presenting a method and code for the calculation of the two-point 
correlation function on the GPU using CUDA. 
We take a different approach to the implementation of the algorithm.  
We have also become aware of earlier work by~\cite{Roeh}, which also takes another different approach to the implementation, and also considers MPI implementation and the use of multiple GPUs in parallel. 

Dark matter cannot yet be detected directly, but the effect of
its gravitational field can be measured indirectly using gravitational
lensing. 
When light travels through the universe, its path is deflected
by the gravitational potential of the matter it passes. 
The distortion of the observed shapes of distant galaxies as their light passes through matter in the universe encodes information about large-scale structure and the growth of matter in the universe. 
However, since galaxies have intrinsic shapes, it is only through statistical analysis of large numbers of galaxies that we can average out their intrinsic shapes and orientations and extract the cosmological information. 
There are many ways to interpret this information (see~\cite{bs} for a review), but for the purposes of this paper we concentrate on the shear peak statistic. 
Background galaxies tend to have the major axis tangential to foreground mass density contours, so concentrations of matter along the line of sight can be detected from the coherent distortions in the ensemble of their shapes. 
Over-densities will appear as peaks in a map of aperture mass. 
Counting the number of peaks as a function of peak significance allows us to distinguish between different cosmologies (\cite{marian09, kratochvil09, dh09}). 
We calculate the aperture mass at a point by making a weighted sum over the tangential components of the ellipticity of the surrounding galaxies. 
Typically this sum contains tens or hundreds of thousands of galaxies, and must be performed on a dense grid of points over the sky in order to accurately reconstruct the projected mass field. 
The sum for each point is independent, and can be performed in parallel making it another good candidate for implementation on the GPU. 
Recent work by ~\cite{leonard} has used wavelet transformations to speed the calculation of the aperture mass, but as far as we are aware this is the first implementation on the GPU. 

This paper is structured as follows. 
In Section~\ref{sec:ang} we introduce the two-point angular correlation function and describe its implementation on the GPU.
We then compare its performance to the same calculation on the CPU using data taken from dark matter simulations. 
In Section~\ref{sec:sps} we describe the aperture mass calculation and how it is implemented on the GPU and compare 
its performance to the CPU implementation. In Section~\ref{sec:summary} we summarize our findings. 
Access to the code is given in \ref{app:coderep}.

\section{Two-point angular correlation function}
\label{sec:ang}

The combined forces of gravitational attraction and dark energy influenced 
the clumping of dark matter, and accompanying galaxy distributions, that
we observe today. To quantify this clumping or clustering, one can make use
of the angular correlation function $w(\theta)$, which relates the probability
$\delta P$ of finding
two galaxies at an angular separation $\theta$ on the observable sky to the 
probability for a random distribution of galaxies (\cite{Peebles1980}):
\begin{equation}
    \delta P = N^2 [1+w(\theta)] d \Omega_1 d\Omega_2,
\end{equation}
where $d\Omega_1$ and $d\Omega_2$ are elements of solid angle and $N$ is
the mean surface density of objects. The full calculation for the angular separation
for any two objects in the sky is shown in \ref{app:asc}.

The estimator for the angular correlation function $\hat{w}(\theta)$ is 
calculated from the positions of galaxies in the sky. Three quantities
go into this calculation: $DD$ (data-data), $RR$ (random-random), 
and $DR$ (data-random). For an angular separation
$\theta$, $DD$ is the number of pairs of galaxies in data separated by $\theta$. 
A random distribution of galaxies is generated over the same region of 
the sky as the data from which $DD$ is calculated, and this random sample is
used to calculate $RR$ in the same fashion as $DD$. $DR$ is the cross-correlation
between these two datasets, i.e., the number of galaxies in the random distribution that
are an angular distance $\theta$ from the galaxies in the data sample. 
While there are different estimators for $w(\theta)$, we
have chosen to work with a widely accepted estimator from \cite{1993ApJ...412...64L},
though we show that the code presented in this paper allows the user to easily 
experiment with other estimators. 
The estimator $\hat{w}(\theta)$ is calculated as
\begin{equation}
    \hat{w}(\theta) = \frac{DD-2DR+RR}{RR}.
\end{equation}

To calculate $DD$, the analyst must calculate
the angular separation from every galaxy to every other galaxy, taking care to 
not calculate the separation of a galaxy from itself or to double count
by calculating the distance from galaxy$_i$ to galaxy$_j$ and then from galaxy$_j$ to galaxy$_i$. 
Given a catalog of $n$ galaxies, there are $n(n-1)/2$ calculations.
There is an analogous number of calculations for $n_r$ randomly distributed
galaxies in the calculation of $RR$. To calculate the $DR$ cross-correlation term,
there are $nn_r$ calculations as every observed galaxy must be correlated
with every simulated galaxy.

The full calculation of $DD$, $RR$, and $DR$ is effectively an $\mathcal{O}(n^2)$ 
operation. 
A standard technique for speeding up the calculation is to use a tree algorithm to make approximations for the density of galaxies at larger angular separations \citep{Jarvis:2003wq}. 
In this kind of technique, the dataset is binned in regions on the sky, and the average of galaxies in that bin is used in place of the individual galaxy values. 
This reduces the calculation time for the two-point function from $\mathcal{O}(n1*n_2)$ to $\mathcal{O}(n_1 + n_2)$, where $n_1$ and $n_2$ are the number of galaxies in two bins. 
If the angular correlation function is to be calculated for bins at a separation smaller than some threshold value $b$ (usually the bin size), then the bins can be split in half and the averages re-calculated. 
Splitting can continue until each bin contains one galaxy, in which case we recover the exact calculation. 
We show here how the exact calculation can be implemented on the GPU for all angular scales, with a similar improvement in calculation time.

\subsection{GPU Implementation}
For the GPU implementation, we make use of the CUDA programming language (\cite{cuda}),
a widely-used language developed by NVIDIA for use with their GPUs. 
In CUDA terminology, threads are the individual calculations performed in parallel by the GPU, and a kernel is the code that defines the calculation and specifies the number of threads that will run this calculation. 
Threads are grouped into blocks, which defines how the threads will be distributed on the GPU, and what sections of local memory they can access. 
To calculate the angular separations that go into $DD$, 
we have noted that there are $n(n-1)/2$ calculations when double counting and self-separations are avoided. 

To calculate DD, we accumulate counts of galaxy pairs in bins of the distance between the galaxies, 
using a predetermined range of angular separation for the binning.
Histogramming on GPUs is nontrivial and NVIDIA provides examples of how 
this can be done (\cite{cudahist}), but we found these difficult to adapt 
to our routines. Therefore we implemented our own solution, which, while 
slower than the highly optimized NVIDIA code, is very flexible and adaptable.

When constructing the algorithm to perform this calculation, we need to consider the memory usage on the GPU. 
The amount of memory available
to the GPU is usually less than that available to the CPU, but if GPU global memory is used for
the initial transfer of data from the CPU and GPU local memory is used only for 
the histogramming, we can avoid this potential limitation (provided the GPU global memory is large enough to hold the full dataset).
However, execution of GPU code can be dramatically slowed if multiple threads were to attempt to access the same memory address at the same time. 
For this reason, we adopt a very trivial algorithm for breaking
down the full calculation into smaller packets, and at the same time using 
atomic operations for the histogramming.

The coordinates for all the galaxies are copied to the GPU in global memory.
The matrix of the necessary calculations is broken into submatrices. 
We loop over these submatrices and for each one  we launch the same kernel, where we pass that kernel the 
matrix indices that define the location of this submatrix in the larger
matrix of calculations. Each column in the submatrix is calculated
by a different thread. 

Shared memory is allocated for each thread-block for the histogramming. The histogram is simply an 
array of integers used to hold the number of galaxies with a particular angular separation.
We increment the entries of the histogram using the {\tt atomicAdd()} operation in CUDA.
There could still be serialization occurring if multiple threads calculate a similar angular separation,
requiring the results to be put into the same bin and therefore accessing the same memory address. 
This will depend on the dataset and the binning used, and can range from a 5\%\ slow-down for fine bins to a 50\%\ slow-down for coarse bins. 
After all the distances are calculated and recorded in the shared memory histogram they 
are summed over all the blocks and copied over to global memory. At this point, the kernel 
returns to the CPU, where the histogram is copied to the GPU memory and added to a histogram
in the CPU memory. 
The full calculation is complete when we have walked through all of the submatrices. 
A summary of the process to calculate $DD$, $RR$, or $DR$ can be stated as follows.

\begin{verbatim}
1. Copy vectors of galaxy positions to 
   the GPU global memory.
2. Determine size and position of 
   submatrices of the calculation.
3. Launch kernel for each submatrix:
     Allocate local memory for 
     histogramming on each block. 
     Each thread calculates the angular 
     separation for a column of the 
     submatrix.
4. Sum the local memory histogram arrays 
   in global memory.
5  Copy the histogram arrays back to the 
   CPU and sum them.
6. Continue to loop over all the 
   submatrices until the entire 
   calculation is done.
\end{verbatim}

The code allows the user to both set the low and high range limits of the histogram
and choose one of three preset bin-widths: evenly spaced bins, logarithmic
binning, and logarithmic base-10 binning. 
Logarithmic binning allows finer bins to be used at small angular separations, where we have the most data. 
We have found that other groups will use different binning when performing their calculations
and so we have built in this flexibility to make it easier for direct comparisons.
The time it takes to process a dataset can depend on these parameters due to the algorithm we have chosen 
to implement. Specifically, if we select a wide range over which to histogram
the distances, the time may be slightly longer because more calculated distances
need to be dropped into the histogram. If the binning is very coarse,
the time may also increase because of serialization issues in memory access. 

We note that for the DR calculation, there is no double counting or self-distances to worry about.
In this case there are about twice as many calculations to be done, but the procedure for 
breaking up the matrix into submatrices is still the same. We show a visualization of this process in Fig.~2.

\subsection{Performance}

We test the consistency of the calculations and measure the increase in speed 
by comparing our GPU implementation to a straightforward implementation on 
a CPU using the C programming language. The details of both the GPU
and CPU computing hardware and environment we used are given in \ref{app:ce}.

The dataset we used is a subset of 
a dark matter N-body simulation that is subsequently ``decorated" with 
galaxies such that the galaxies realistically trace the underlying dark matter distribution (~\cite{wechsler, busha}). 
We use galaxies from the simulation with a redshift of $z<0.1$.
The dataset is read in from a text file that has two columns: the right ascension (RA)
and declination (DEC) in arc minutes for each galaxy.
The angular range covered in this dataset is $0^\circ<\textrm{RA}<90^\circ$ 
and $0^\circ<\textrm{DEC}<90^\circ$ and 
we generate a flat distribution of galaxies over the same range. 
We use this dataset to demonstrate the speed of the
full calculation for different numbers of galaxies. 

The first step is to show that the GPU implementation gives the same values as
those calculated with the CPU. We compare the two implementations using $10^{5}$
galaxies for the $DR$ component, which requires $n^2=(10^5)^2=10^{10}$ calculations.
We use 254 bins and logarithmic binning with a wide range of histogram bins. 
There are 172 non-empty bins and only 30 of them have exactly the same numbers when 
the CPU or GPU are used. However, closer inspection of bins with different numbers of entries shows they
are almost exactly the same. 
The fractional difference between the bins is always between 
$10^{-3}-10^{-8}$, with the smaller fractional differences found in bins with 
a large number of entries.
We attribute this discrepancy to either differences in floating point implementation 
or trigonometric and logarithmic implementations on  the CPU and GPU.

For timing comparisons using the CPU and GPU, we choose the $DR$ 
calculation for $10^{4}$, $10^{5}$, and $10^{6}$ galaxies. We use 254 bins and logarithmic
binning. We note that to calculate either $DD$ or $RR$ each require about one-half
the number of calculations as $DR$ and so the full calculation for $\hat{w}(\theta)$
should take about twice the amount of time shown here, given only one CPU or GPU.
The times are calculated using the UNIX/Linux command {\tt time}, which gives the total
time required to execute any command and return control to the user. 
While this gives less fine-grained
information about where the programs spend their time, we choose this method
as it gives the best sense of how much time it costs the analyst to actually make 
these calculations. While both the CPU and GPU implementation spend most of their
time actually calculating the angular separations, this is a slightly smaller percentage of the 
total time when using the GPU. Therefore, when testing with small numbers of galaxies,
the GPU implementation does not quite increase as $n_{\rm galaxies}^2$.
The results are summarized in Table~\ref{tab:acf}.
For $10^{5}$ galaxies, we note that there is a 140$\times$ increase in speed when using the GPU
and for $10^{6}$ galaxies a 180$\times$ increase in speed.

\begin{center}
\begin{table}
\caption{Speed of angular correlation calculations (in seconds) when performed on either the CPU or GPU, 
for different numbers of galaxies. The times are for the $DR$ component; there are
$n^2$ calculations performed, where $n$ is the number of galaxies. All times are in seconds.
These tests used a histogram with 254 bins
and logarithmic binning. Compilation details can be found in \ref{app:ce}. \label{tab:acf}}
\begin{tabular}{lccc}
\hline \hline
    & $10^{4}$ galaxies & $10^{5}$ galaxies & $10^{6}$ galaxies\\ 
\hline 
CPU & 23.9 & 2305 & 231,225 \\ 
GPU & 0.2 & 16.0 & 1,254 \\
\hline
\end{tabular}
\end{table}
\end{center}

For completeness we ran a variety of $DR$ calculations with the $10^{5}$ galaxy sample, where we
vary the number of bins, the range of bins, and whether or not we use equal bin widths or
logarithmic binning (natural and base 10). As we mentioned before, if too many threads are accessing
the same histogram bin, then the calculation can take longer. 
In addition, if the range of the bins is less than the range of the data, time can be saved by ignoring calculations that would fall outside the bin range. 
Running over a variety of
these combinations, we found that the $DR$ calculation for $10^{5}$ galaxies can take
between 11 and 57 seconds. Therefore users should be aware that depending on their choice
of binning, timing results may vary. 

We note that the implementation described here does not use the GPU with maximum efficiency. 
Modern GPUs can perform memory transfer and kernel launches in parallel, and we plan to take advantage of this capability in future code development. 
This will be particularly important for use with future, larger datasets. 

\section{Shear Peak Statistics}
\label{sec:sps}
As described in Section~\ref{sec:intro}, light from distant galaxies is lensed by structures along the line-of-sight, and therefore the statistical ensemble of observed galaxy shapes encodes information about the matter power spectrum of the universe. 
This information can be used to constrain models of cosmology, which predict differing amounts of structure on different mass and distance scales, and in different epochs. 

The tidal gravitational field of matter along the line-of-sight causes 
the shear field to be tangentially aligned around projected mass-density peaks. 
This alignment can be used to detect matter over-densities by constructing the aperture mass ($M_{ap}$) described in ~\cite{Schneider} which is the weighted sum of the tangential components of shape of the galaxies ($\epsilon_t$) surrounding a position in the sky ({\boldmath{$\theta_0$}}):
\begin{equation}
    M_{ap} ( \boldmath{\theta_0}) = \frac{1}{n_{gal}} \sum\limits_{i}  Q(\theta_i)\epsilon_{i,t}, 
\end{equation}
If the filter function $Q$ has a shape that follows the expected shear profile of a mass peak, then the aperture mass is a matched filter for detecting these peaks. 
$Q$ can have a generic (e.g. Gaussian) shape, or be optimized for detections of halos with an NFW density profile (\cite{1996ApJ...462..563N, Schirmer}):
\begin{eqnarray}
Q_{NFW}(x) =  \frac{1}{1+e^{6 - 150x} + e^{-47 + 50x}} \frac{\tanh(x/0.15)}{x/0.15},
\label{eqn:mAp-filt}
\end{eqnarray} 
where $x = \theta_i/\theta_{max}$, and $\theta_{max}$ defines the radius of the filter. 
See~\cite{maturi} for a detailed study of different filter shapes and their impact on cosmological constraints. 
The filter radius is chosen to maximize the sensitivity to cosmology, and typically several different filter sizes are combined to obtain information on both large-scale and smaller-scale structures in the matter density of the universe. 
We look for peaks in the map of signal-to-noise ratio (SNR), where the noise can also be calculated directly from the data:
\begin{equation}
{\rm SNR}(\theta_0) = \frac{\sqrt{2}\sum_i Q(\theta_i)\epsilon_i} {\sqrt{ \sum_i Q^2(\theta_i)\epsilon^2}}. 
\end{equation}

The easiest way to implement this algorithm on the CPU is with compiled C code, using nested loops over vectors containing the measured galaxy parameters. 
We can reduce the time required for calculation by  considering only galaxies that have a significant contribution to the aperture mass. 

The complexity of this algorithm is $\propto O(n_{pts} * n_{gals})$, where $n_{pts}$ is the number of points in space for which we wish to evaluate the aperture mass (e.g. a grid of 512$\times$512 points), and $n_{gals}$ is the number of galaxies that significantly contribute to the aperture mass. 
Since we ignore galaxies far from the reconstruction point where the value of the filter function goes to zero, $n_{gals}$ is proportional to the filter radius squared, $\theta_{max}^2$, as given in Equation~\ref{eqn:mAp-filt}.

This becomes a computational problem with a large dataset, since we must make a nontrivial calculation of aperture mass  for all surrounding galaxies for each point on the sky. 
For small area or shallow surveys this is a feasible calculation on a desktop machine, provided we use a low-density grid of reconstruction points. 
Larger, deeper surveys, which contain a higher galaxy density and a larger area over which to calculate the aperture mass, become problematic. 

\subsection{GPU Implementation}
\label{subsec:mAp-gpu}
 
The implementation on the GPU is straightforward, and simply parallelizes the brute-force method described above, using CUDA. 
First, the vectors of galaxy parameters are copied into global memory on the GPU. 
The kernel is then launched, where each thread calculates the contributions of the surrounding galaxies for one point on the reconstruction grid, summing the contributions and returning the resulting aperture mass and SNR. 
The number of threads required is equal to the number of points in the reconstruction grid. 
The process can be broken down as follows: 
\begin{verbatim}
1. Copy vectors of galaxy positions and 
   shear components to the GPU global 
   memory.
2. Launch kernel:
     Each thread calculates the aperture 
     mass for one point on the grid, 
     looping over the surrounding 
     galaxies and summing the 
     contributions to the aperture mass. 
3. Copy vector of aperture mass from GPU 
   global memory. 
\end{verbatim}

The complexity of this algorithm is $ O(n_{gals})$, or $ O(\theta_{max}^2)$, since we parallelize the calculations over the grid of reconstruction points. 

GPU units designed for scientific computing have a large on-chip memory. 
We use a Tesla M2070 GPU card with Fermi architecture and 5.25GB on-chip global memory (with ECC on). 
See \ref{app:ce} for more details on the system architecture. 
This is sufficient memory to store the positions and shear components for tens of millions of galaxies. 
The more important limit in the amount of data we can process in a single kernel launched on the GPU is the number of threads that can be launched simultaneously.  
This will eventually limit the possible density of the reconstructed aperture mass map. 
By splitting the sky into overlapping segments and calculating the aperture mass for each segment sequentially we can easily circumvent this issue. 
However, for the survey size and reconstruction grid density we consider in this work, it has not been necessary to do this. 
All performance numbers in Section~\ref{subsec:mAp-perf} are given for all data processed in a single kernel launch with no segmentation of the dataset required.

\subsection{Performance}
\label{subsec:mAp-perf}
\begin{figure}
\includegraphics[width=0.5\textwidth]{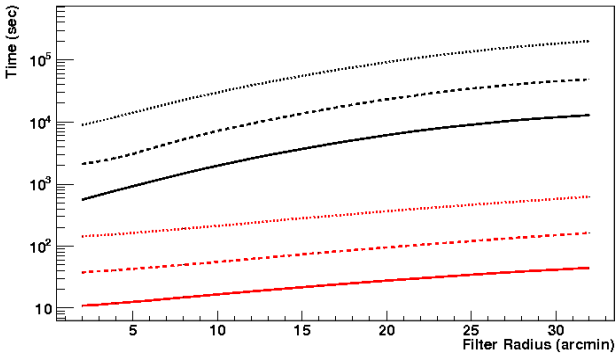}
\caption{Calculation time for CPU code (top, black) and GPU code (bottom, red) for increasing filter size. The solid lines correspond to a reconstruction grid of 512$\times$512 points, dashed lines are for 1024$\times$1024 points and dotted lines for a grid of 2048$\times$2048 points. }
\label{fig:mAp}
\end{figure}

We compare the performance of the CPU and GPU algorithms to analyze realistic shear data. 
We create input catalogs from simulated shear fields created from a ray-traced N-body dark matter simulation from ~\cite{kratochvil09}, with model galaxies as tracers of the shear field. 
We use  a galaxy density of 35 galaxies per arcmin$^2$ (which is roughly the galaxy density expected to be used in LSST weak lensing analyses (~\cite{lsst-sb}). 
There are a total of $10^6$ galaxies in our dataset, which represents an area of sky of $\sim$ 7.9 deg$^2$. 
Using more or less galaxies does not impact the computational time, only the memory usage and the time required for transferring data to the GPU, so we compare compute times for different filter sizes and different densities of the reconstruction grid. 
Table~\ref{tab:mAp} shows the speed for the CPU and GPU implementation of the aperture mass code in these different cases. 
This information is shown graphically in Figure~\ref{fig:mAp}. 
We can see from this figure that, for a constant dataset and grid size, the calculation speed scales as the square of the filter size for the GPU calculation. 
The scaling is different for the CPU calculation, because the optimisation flags we use to compile the code allow loop unrolling (see ~\ref{app:ce} for details) which becomes more significant at larger filter radius. 

As for the timing tests for the angular correlation function, we include in our GPU timing the time required for all data to be copied to and from the GPU global memory. 
This comprises $\sim$ 3 seconds and is constant for all filter and grid sizes. 
The aperture mass maps obtained from both methods are identical. 

\begin{table*}[htb]
\begin{center}
\caption{Speed of calculation (in seconds) for CPU and GPU code for increasing density of reconstruction grid, and for increasing filter radius. The GPU code gives calculation times  100 - 300 times faster. 
\label{tab:mAp}}
\begin{tabular}{rccccc}\hline \hline
& $2'$  & $4'$  & $8'$ & $16'$ & $32'$ \\ \hline
\multicolumn{6}{c} {CPU} \\
512$\times$512 grid   & 558 & 781 & 1,478 & 4,108 & 12,794 \\ 
1024$\times$1024 grid  & 2,105 & 2,627 & 5,204 & 15,311 & 48,474 \\ 
2048$\times$2048 grid  & 8,855 & 11,941 & 22,546 & 61,406  & 189,861 \\ \hline

\multicolumn{6}{c} {GPU} \\
512$\times$512 grid  & 11 & 12 & 15 & 23 & 45 \\  
1024$\times$1024 grid   & 38 & 41 & 50 & 78 & 162 \\ 
2048$\times$2048 grid  & 143 & 156 & 190 & 297 & 627  \\ \hline
\end{tabular}
\end{center}
\end{table*}

Using the GPU code decreases the calculation time by 100$\times$-300$\times$  compared to the calculation on the CPU.  
Recent work by \cite{leonard} has produced a very fast method of calculating the aperture mass using a wavelet transformation. 
They achieved speed-up factors from 5x-1200x, depending on filter radius, for an image of 1024x1024 pixels. 
We take a very different approach to the aperture mass calculation, but we find similar improvements in speed, with the advantage that our approach calculates the noise directly from the data (as outlined in \cite{bs}), whereas the \cite{leonard} technique requires randomization of the data and repeated measurements to obtain the uncertainty. 
Our approach saves considerable computing time. 
The timings given in this section include the time necessary for calculation of the noise.

\section{Summary}

\label{sec:summary}
We have implemented code to perform calculations of the two-point angular correlation function and the aperture mass statistic on the GPU. 
We have demonstrated that this implementation can reduce compute times for these calculations by factors of 100$\times$-300$\times$, depending on the amount of data to be processed. 
The code for making this calculation is publicly available from Github (see \ref{app:coderep} for details). 
With a straightforward division of the dataset into sub-sets, our code can also be used on clusters of independent GPUs. 
Faster compute speeds mean that a full MC-based calculation of the errors for the angular correlation function can reasonably be performed, without approximations or assumptions that are required to make the calculation reasonable for CPU codes (e.g. kd-tree calculation (\cite{Jarvis:2003wq})). 
We intend to evaluate this in future work. 

The increasing size of astronomical datasets will require a new approach to data analysis. 
We expect that the use of the GPU in everyday cosmological calculations will become more common in the next few years, especially since faster compute times allows experimentation in techniques used to make the calculation and rapid comparison of the results. 
We expect this application to be extended to other computationally challenging calculations, such as the three-point and higher order angular correlation functions, and the shear correlation functions.


\appendix

\section{Angular separation calculation \label{app:asc}}

The angular distance between two galaxies can be calculated using:

\begin{eqnarray*}
    \Delta \alpha &=& \alpha_2 - \alpha_1 \\
                A &=& \cos^2\delta_2\sin^2\Delta \alpha\\
                B &=&  \cos\delta_1\sin\delta_2 - \sin\delta_1\cos\delta_2\cos\Delta \alpha\\
                C &=&  \sin \delta_1 \sin\delta_2 + \cos\delta_1\cos\delta_2 \cos\Delta \alpha \\
           \theta &=& \frac{180^\circ}{\pi}\arctan \left( \frac{\sqrt{A+B^2}}{C} \right)
\end{eqnarray*}

where $\theta$ is the angular separation between any two galaxies in degrees and 
($\alpha_i,\delta_i$) is the right ascension and declination of the $i^{\rm th}$ galaxy.

\section{Computing environment \label{app:ce}}

The GPU implementations for both algorithms were tested on a 
Tesla M2070. This device has 5375 MB of global memory (with ECC turned on) 
and 49152 bytes of shared memory. 
The clock speed for the GPU processors is 1.15 GHz.
For tests using the CPU, we run on an 
Intel Xeon E5540 processor with a 2.53 GHz
clock speed and 3 GB of on-board memory.

We use floating-point precision for all our calculations; we found that using double 
precision had no impact on our results. 

We compile our code using CUDA version 4.10 (both driver and runtime)
and {\tt gcc} version 4.1.2. The operating system is Scientific Linux,
version SL5, using kernel 2.6.18.

For the timing comparisons
, we write the
CPU code using C/C++ and compile
using the {\tt gcc} compiler with the optimization flags {\tt -O1}.
This flag gives about a 30\% improvement over not using this flag. 
Greater degrees of optimization ({\tt -O2, -O3}) did not give any additional
increase in speed.

\section{Code repository \label{app:coderep}}

Our code is publicly available on Github, a software hosting service that uses
{\tt git} for version control. The repository can be found at

\noindent{\tt https://github.com/djbard/ccogs} and can be cloned by anyone who
has {\tt git} installed on their system.

Along with the code, we have provided sample datasets and scripts to run
and test your installation. Each package has its own {\tt README} that 
details how to build and run the software. Problems or improvements can 
be directed to the authors.

This software is licensed under the MIT License.

\bibliographystyle{model2-names}
\bibliography{paper}

\end{document}